\documentclass{ws-ijmpc}
\usepackage{graphicx}
\usepackage{dcolumn}
\usepackage{bm}
\usepackage{epsfig}
\usepackage{color}

\begin{document}
\title{High-Order Kinetic Relaxation Schemes \\ as High-Accuracy Poisson Solvers}

\author{M. Mendoza} \address{ ETH Z\"urich, Computational Physics for
  Engineering Materials, Institute for Building Materials,
  Wolfgang-Pauli-Str. 27, HIT, CH-8093 Z\"urich, Switzerland. \\
  mmendoza@ethz.ch}

\author{S. Succi} \address{Istituto per le Applicazioni del Calcolo
  C.N.R., Via dei Taurini, 19 00185, Rome, Italy. \\ succi@iac.cnr.it}

\author{H. J. Herrmann}\address{ ETH Z\"urich, Computational Physics
  for Engineering Materials, Institute for Building Materials,
  Schafmattstrasse 6, HIF, CH-8093 Z\"urich, Switzerland. \\
  hjherrmann@ethz.ch}

\maketitle


\begin{abstract}
We present a new approach to find accurate solutions to the Poisson equation, as obtained from the steady-state limit of a diffusion equation with strong source terms. For this purpose, we start from Boltzmann's kinetic theory and investigate the influence of higher order terms on the resulting macroscopic equations. By performing an appropriate expansion of the equilibrium distribution, we provide a method to remove the unnecessary terms up to a desired order and show that it is possible to find, with high level of accuracy, the steady-state solution of the diffusion equation for sizeable Knudsen numbers. In order to test our kinetic approach, we discretise the Boltzmann equation and solve the Poisson equation, spending up to six order of magnitude less computational time for a given precision than standard lattice Boltzmann methods. 
\end{abstract}


\maketitle

\section{Introduction}

The diffusion equation is widely used to describe transport phenomena 
in which heat, mass, and other physical quantities are transferred in space and time 
due to underlying molecular collision processes \cite{ottinger2005}. At steady-state and when the source term does not depend explicitly on the concentration of the associated field, such diffusion processes settle down in the form of a non-local relation between the spatial distribution of the source and the resulting concentration of the associated field, a relation which is governed by the Poisson equation \cite{poisson_acc}. The notion of the Poisson equation as the steady-state solution of the diffusion  equation extends to many other phenomena, e.g. electrostatics systems \cite{electro}, plasma physics \cite{ottaviani}, chemistry and biology \cite{chemi-solutions,chemi-bio}, molecular biophysics \cite{molecular-bio}, density functional theory and solid-state physics \cite{EGROSS,kaxiras1,DFTbook}, to name but a few. In fact, solving the Poisson equation accurately and efficiently continues to be a subject of intense research to this day.

From a microscopic point of view, the diffusion equation emerges from the underlying Boltzmann kinetic equation in the limit of very small Knudsen numbers, i.e. molecular mean free path much shorter than  the typical macroscopic scale. Solving the Boltzmann kinetic equation to study diffusion processes makes apparently little sense, since the Boltzmann distribution function lives in a double-dimensional phase-space, i.e. position and velocity. However, in the last two decades, minimal (lattice) versions of the Boltzmann equation have been developed, in which velocity space is reduced to a small set of discrete velocities, so that the computational cost is cut down dramatically, making the solution of the lattice kinetic equation often competitive towards standard grid-discretisation of the corresponding partial differential equation.

The Lattice Boltzmann Method (LBM) has attracted much interest for solving the Navier-Stokes equations \cite{LBE1,LBE2}, and has been extended to solve other type of phenomena, e.g. diffusion equation \cite{diffusion,diffusion2}, Maxwell equations\cite{electroLB}, quantum systems \cite{QLB}, relativistic fluid dynamics \cite{rlbPRL,rlbPRD}, and the Poisson equation \cite{poisson_acc,poisson_1,poisson_2,poisson_3,poisson_4}, to name but a few. To the best of our knowledge, to date, the solution of the Poisson equation based on kinetic theory has been confined to smooth (non-stiff) source terms and small Knudsen numbers \cite{diffusion-source}. If the target is the steady-state solution of the diffusion equation, as it is the case for the Poisson equation, the kinetic approach looses much of its appeal, since its inherently real-time dynamics must proceed through smaller time-steps than those affordable by fictitious-time iteration techniques. In this paper we shall show that such gap can be closed by resorting to higher-order kinetic formulations, at least for the cases where the source term does not depend on the concentration field avoiding artificial solutions as pointed out in Ref.~\cite{poisson_acc}. 

Higher-order formulations of the LB equation have been developed before, mostly in the context of thermal \cite{karli} and relativistic fluid dynamics \cite{rlbDIS}, and recently, density functional theory \cite{dftLB}. There, the main idea is to design the equilibrium distribution in such a way to include not only the conserved moments, such as density and current, but also the non-conserved ones, namely the momentum flux tensor and the heat flux, and eventually higher order kinetic moments with no direct hydrodynamic significance (sometimes called ``ghosts''). Clearly, in order to match this longer ladder of kinetic moments, a correspondingly larger set of discrete velocities is required. It should be emphasized that the discrete kinetic equation is a superset of the diffusion equation, and consequently, to the purpose of solving the diffusion equation alone, higher-order contributions must be regarded as discretisation artefacts which need to be minimised, and possibly canceled altogether. This is precisely the target of this paper. Here, we will use higher-order lattices and associated equilibria, to remove higher-order contributions to the  macroscopic equations reproduced by the Boltzmann equation. For this purpose, we develop the respective LBM and show that, by expanding the equilibrium distribution up to seventh order in Hermite polynomials, one can solve the Poisson equation {\it six orders} of magnitude faster than with standard LB models, at a given level accuracy. 

The paper is divided as follows: In Sec.~\ref{sec1}, we introduce the diffusion equation with stiff source  term and the associated Boltzmann kinetic equation. 
In Sec.~\ref{sec2}, we describe the influence of higher order moments on the steady-state solution of the diffusion equation. In Sec.~\ref{sec3}, we develop a LBM for the Poisson equation, validate and compare the model with standard cases. Finally, we discuss the results and outlook of our work. 

\section{From Boltzmann kinetic theory to diffusion equation}\label{sec1}

Let us begin by writing the diffusion equation for a scalar $\Phi$ in the standard form: 
\begin{equation}
\label{DIFFU}
\frac{\partial \Phi}{\partial t} = D \nabla^2 \Phi  + S \quad ,
\end{equation}
where $\Phi$ is the concentration of particles (temperature in the case of the heat transfer equation), $D$ is the diffusivity, and $S$ is the source term, which, in our case, depending neither on
$\Phi$ nor on time. 
The steady-state version of Eq.~\eqref{DIFFU} delivers the Poisson equation with source $-S/D$. It is well-known that this equation can be obtained from an underlying Boltzmann kinetic equation, via a Chapman-Enskog expansion, in the limit of small Knudsen numbers, $Kn = \lambda/L$, being $\lambda$ the mean-free path and $L$ the characteristic size of the system \cite{chapman}. 

The Boltzmann equation in the single relaxation time approximation (BGK) \cite{BGK} reads as follows:
 \begin{equation}\label{boltzmann}
   \frac{\partial f}{\partial t} + \vec{v}\cdot \nabla f = -\frac{1}{\tau} (f - f^{\rm eq}) + {\cal S}\quad ,
 \end{equation}
where $f\equiv f(\vec{x}, \vec{v}, t)$ is the probability distribution function, $\vec{v}$ are the microscopic velocity vectors, $\tau$ is the relaxation time, and $f^{\rm eq}$ is the equilibrium distribution. 

In the non-relativistic context, the local equilibrium is given by a Maxwell-Boltzmann distribution:
\begin{equation}
f^{eq} = \frac{e^{-{(v-u)}^2/2 c_s^2}}{(2\pi c_s^2)^{3/2}}  \quad ,
\end{equation}
where $u$ is the local flow speed. Such local equilibrium encodes the basic symmetries of Newtonian mechanics, particularly Galilean invariance, which is built-in via the dependence on the molecular speed relative to the fluid $v-u$, rather than the absolute one, $v$. In order to secure Galilean invariance for {\it any} fluid speed, kinetic moments at {\it all orders} should be matched on the lattice. This would require an infinite connectivity, which is clearly unviable on any realistic lattice. As a result, LB local equilibria are typically based on finite-order Hermite expansions, of the form 
\cite{Grad}:
\begin{equation}
  f^{\rm eq}(\vec{x}, \vec{v}, t) = w(\vec{v}) \sum_{n=0}^{\infty} a_n^{(n)}(\vec{x}, t) H_n^{(n)}(\vec{v}) \quad ,
\end{equation}
where $H_n^{(n)}$ are the tensorial Hermite polynomials of order $n$, and $w(v)$ is the weight function defined as,
\begin{equation}
  w(\vec{v}) = \frac{1}{(2\pi c_s^2)^{3/2}} e^{-v^2/2 c_s^2} \quad ,
\end{equation}
$c_s$ being the sound speed. Note that $w(v)$ is the uniform {\it global} equilibrium, corresponding the no-flow limit of the local equilibrium $f^{\rm eq}$. The coefficients $a_n^{(n)}$ (kinetic moments) are calculated by projecting the equilibrium distribution onto the respective Hermite polynomial,
\begin{equation}\label{coefficients}
a_n^{(n)} = \int f^{\rm eq} H_n^{(n)}\; d^3 v \quad .
\end{equation}
The lowest-order kinetic moments have a direct macroscopic interpretation.
For instance, by integrating on the velocity space ($H_0=1$),
\begin{equation}
\Phi = \int f \; d^3 v = \int f^{\rm eq} \; d^3 v \quad ,
\end{equation}
where the last equality comes from the requirement of mass conservation 
on the collsion operator, namely:
\begin{equation}
\frac{1}{\tau} \int (f - f^{\rm eq} ) \; d^3 v = 0 \quad .
\end{equation}
The diffusivity of the system is dictated by the relaxation time as 
\begin{equation}
D = c_s^2 \tau.
\end{equation}

It can be shown that the Boltzmann equation recovers the diffusion equation in the 
limit of small Knudsen numbers, which is related with the mean-free path and therefore 
with the relaxation time. This is the strongly-interacting fluid regime, whereby the particle dynamics is
collision-dominated, so that collective behaviour sets in on a short timescale $\tau$.
The opposite limit of large Knudsen numbers, hence large values of $\tau$, corresponds to the free-particle motion (ballistic regime), whereby memory of the initial conditions is kept for a very long time. It is therefore clear that in the ballistic regime higher order moments of the equilibrium distribution do not relax and consequently the Boltzmann equation does not converge to any standard diffusion equation. On the other hand, by increasing the relaxation time, hence the effective diffusivity, one would intuitively expect a quickest path to steady-state. 
In this respect, it would be highly desirable to derive a diffusion equation in kinetic 
form, which can attain steady-state solution at relatively large Knudsen numbers by
exactly cancelling as many high-order terms as possible.

As discussed above, this can be achieved by zeroing as many higher order contributions as possible, while retaining the largest possible diffusion coefficient.  

\section{Influence of Higher-Order Terms}\label{sec2}

Let us consider the case when all time derivatives are zero, $\partial /\partial t = 0$. 
The Boltzmann equation, Eq.~\eqref{boltzmann}, can be written as,
 \begin{equation}\label{boltzmann1}
 f  = f^{\rm eq} + \tau {\cal S} -\tau \vec{v}\cdot \nabla f \quad .
 \end{equation}
By integrating this equation in velocity space, we obtain
a steady-state continuity equation for the current density:
\begin{equation}\label{first}
\nabla \cdot \vec{J} = S \quad ,
\end{equation}
where $\vec{J} = \int f \vec{v}\; d^3v$ and $S = \int {\cal S}\; d^3 v$. 

Multiplying Eq.~\eqref{boltzmann1} by $\vec{v}$ and integrating again, we obtain 
 \begin{equation}\label{second}
   \vec{J} = \vec{J}^{\rm eq} + \tau \vec{S} - \tau \nabla \cdot \Pi \quad ,
 \end{equation}
with $\Pi \equiv \Pi_{\alpha \beta} = \int f v_\alpha v_\beta \; d^3v$ being the
momentum-flux tensor and $\vec{S} = \int {\cal S} \vec{v}\; d^3v$ the source current.  
The first term at the right-hand-side takes the form 
$\vec{J}^{eq} = \Phi \vec{u}$, where $\vec{u}$ is the flow speed in the local equilibrium.
For the case of a purely diffusive dynamics this term is zero.
The second term contributes a shift $\tau \vec{S}$ to the flow speed, and it must also be
set to zero for the case of pure diffusion.
Finally, the pressure tensor shoud reduce to $c_s^2 \Phi$, so that inserting (\ref{second})
into (\ref{first}), the diffusion equation is obtained.
The first two conditions are automatically ensured by setting $\vec{u}=0$ in the local equilibrium.
The third one, however, cannot be enforced exactly because of higher order contributions to the
momemtum flux tensors.

Indeed, by iterating the procedure, one obtains the general expression:
\begin{equation}\label{master:eq}
\sum_{n=0}^\infty (-1)^n \tau^n \nabla^{n+1} \left [ \Pi^{{\rm eq} (n+1)} + \tau S^{(n+1)} \right] = S \quad ,
\end{equation}
where $\Pi^{{\rm eq}(n)} \equiv \int f^{\rm eq} v_{\alpha_1} ... v_{\alpha_n} \; d^3 v$ and $S^{(n)} \equiv\int {\cal S} v_{\alpha_1} ... v_{\alpha_n} \; d^3 v$, are both tensors of order $n$. Note that for small Knudsen numbers ($\tau \nabla \ll 1$)  
higher-order terms become negligible and the series is convergent. We also observe that the solution is dictated by the equilibrium distribution and the source term, so that, by properly choosing both expressions, higher order terms can be canceled thus opening the way to high-accuracy solutions of the Poisson equation.
 
In numerical practice, accuracy is typically improved by increasing the resolution of the grid.  This imposes correspondingly smaller time steps, which is clearly expensive, especially in three dimensions. High-order methods are meant to mitigate the problem by achieving the same level of accuracy with many less grid points. They are typically based on higher-order stencils for the discretisation of the corresponding differential operators.

In this work, we present a procedure whereby the same goal is achieved by drawing directly from an underlying lattice kinetic theory, i.e. by a suitable (non-Gaussian) extension of the local kinetic equilibria as combined with the use of larger sets of discrete speeds than those typically employed in standard Lattice Boltzmann theory.
 
\section{Kinetic approach to the Poisson equation}\label{sec3}

In order for the Boltzmann kinetics to converge to the Poisson equation, the following conditions need to be met: 
\begin{eqnarray}
\label{CON}
\Pi^{{\rm eq}(1)} = \Pi^{{\rm eq}(n > 2)} = 0,\\
S^{(n>0)} = 0 \nonumber. 
\end{eqnarray}
The equilibrium distribution can be expressed as a separable product of 
a function of the microscopic velocity (global uniform equilibrium) and 
a function of the time-spatial coordinates, namely: 
\begin{equation}\label{equilibrium2}
f^{\rm eq} = \phi (\vec{v})\Phi (\vec{x}, t) \quad ,
\end{equation}
and similarly for the source term:
\begin{equation}
  {\cal S} = \chi (\vec{v}) S(\vec{x}, t) \quad .
\end{equation}
In the above, $\chi(\vec{v})$ and $\phi(\vec{v})$ are velocity dependent functions that need to be determined based on the conditions \eqref{CON} above. To this purpose, we write:
\begin{equation}
  \phi(\vec{v}) = w(\vec{v}) \sum_{n=0}^{N} b_n^{(n)} H_n^{(n)}(\vec{v}) \quad ,
\end{equation}
and,
\begin{equation}
  \chi (\vec{v}) = w(\vec{v}) \sum_{n=0}^{N} c_n^{(n)} H_n^{(n)}(\vec{v}) \quad ,
\end{equation}
where $b_n^{(n)}$ and $c_n^{(n)}$ are constant coefficients and $N$ is a 
cut-off truncating the Hermite expansion. With reference to the one-dimensional case and an expansion up to seventh order ($N=7$), we obtain:
\begin{equation}\label{omega:eq}
  \phi(v) \simeq w(v)\left ( 1 - \frac{3c_s^2 - 6 c_s^2 v^2 + v^4}{8 c_s^4} - \frac{15 c_s^6 - 45 c_s^4 v^2 + 15 c_s^2 v^4 - v^6}{24 c_s^6}\right ) \quad ,
\end{equation}
and 
\begin{equation}\label{chi:eq}
  \chi(v) \simeq w(v)\left ( 1 + \frac{1}{2}\left [ 1 - \frac{v^2}{c_s^2}\right ] + \frac{3c_s^2 - 6 c_s^2 v^2 + v^4}{8 c_s^4} + \frac{15 c_s^6 - 45 c_s^4 v^2 + 15 c_s^2 v^4 - v^6}{48 c_s^6}\right ),
\end{equation}
where the weight $w(v)$ for the one-dimensional case is defined as
\begin{equation}
w(v) = \frac{1}{(2\pi c_s^2)^{1/2}} e^{-v^2/2 c_s^2} \quad .
\end{equation}
With Eqs.~\eqref{omega:eq} and \eqref{chi:eq}, one can prove that the
the following expressions for the moments are fulfilled: $\Pi^{{\rm eq}(0)}=\Phi$, $\Pi^{{\rm eq}(1)}=0$, $\Pi^{{\rm eq}(2)}=\Phi c_s^2$, $\Pi^{{\rm eq}(n>2)}=0$, $S^{(0)} = S$, $S^{(n>0)} = 0$. 
These expressions hold only up to $N = 7$. 

To develop a lattice Boltzmann model and solve the Poisson equation numerically, we need to impose a quadrature and find a corresponding set of finite velocity vectors, $v_i$, such that the orthogonality conditions are fulfilled to the desired order \cite{karli,rlbDIS}, namely:
\begin{equation}\label{quadrature}
\int w(v) H_n (v) H_m (v)\; dv = \sum_{i = 0}^{N_v} w_i H_n (v_i) H_m (v_i) \quad . 
\end{equation}
Here $w_i$ are discrete weights and $H_n$ are the one-dimensional Hermite polynomials. 
The number of velocity vectors $N_v$ depends on the order of the expansion that we intend to reproduce 
with the quadrature. For expansions up to seventh order, we need at least $13$ velocity 
vectors and weights. Each set of numbers also provides its own lattice sound 
speed $c_s$, which goes from about $0.4$ with $N=3$ to nearly $1.2$ with $N=7$. 
The details of these values are given in \ref{app}.

Higher-order schemes are usually exposed to numerical instabilities, due to the appearance of additional modes in the dispersion relation. However, since our procedure is designed to annihilate precisely these modes, we expect our approach to be stable. Furthermore, the method proposed here can also be used as a systematic technique
of calculating weights and velocity vectors, such that one can achieve the
desired level of accuracy for the computation of Laplacian operators, similar 
to the work proposed in Refs. \cite{ansumali1,ansumali2}. 

We have performed the theoretical analysis in one-dimension, however, extensions to two- and three-dimensions are straightforward, by using the tensorial form of the Hermite polynomials and by finding the discrete velocity vectors with the respective orthogonality conditions. For instance, in order to fulfill the conditions \eqref{CON}, up to fifth order, the following expansions apply:
\begin{equation}\label{omega:eq3}
  \phi(\vec{v}) \simeq w(\vec{v}) \left ( 1 - \frac{15 c_s^4 - 10 c_s^2 \vec{v}^2 + \vec{v}^4}{8 c_s^4} \right )\quad ,
\end{equation}
and 
\begin{equation}\label{chi:eq3}
  \chi(\vec{v}) \simeq w(\vec{v})\left ( 1 + \frac{1}{2}\left [ 3 - \frac{\vec{v}^2}{c_s^2}\right ]  + \frac{15 c_s^4 - 10 c_s^2 \vec{v}^2 + \vec{v}^4}{8 c_s^4} \right ),
\end{equation}
where the weight $w(\vec{v})$ is defined as
\begin{equation}
w(\vec{v}) = \frac{1}{(2\pi c_s^2)^{3/2}} e^{-\vec{v}^2/2 c_s^2} \quad .
\end{equation}
For the three-dimensional case, we should calculate the corresponding set of discrete velocity vectors, $\vec{v}_i$, such that the orthogonality condition is fulfilled up to fifth order, namely:
\begin{equation}\label{quad3}
  \int w(\vec{v}) H_m^{(m)}(\vec{v})  H_n^{(n)}(\vec{v}) d^3v = \sum_{i=0}^{N_v} w_i H_m^{(m)}(\vec{v}_i)  H_n^{(n)}(\vec{v}_i) ,
\end{equation}
which is the three-dimensional version of Eq.~\eqref{quadrature}. By solving these algebraic equations, we find that we need at least $111$ velocity vectors and $10$ weights. Details are given in \ref{app}. 

\section{Numerical Results}

As an application, we solve the Poisson equation \cite{electro},
\begin{equation}\label{poisson1}
  \nabla^2 \Phi = -\frac{\rho}{\epsilon} \quad ,
\end{equation}
where $\rho$ is the charge density and $\epsilon$ is the electric permittivity. 

Replacing the equilibrium distribution $f^{\rm eq}$ from Eq.~\eqref{equilibrium2} 
into Eq.~\eqref{master:eq}, we obtain:
\begin{equation}
S + c_s^2 \tau \nabla^2 \Phi \simeq 0 \quad ,
\end{equation}
implying that $S=\rho c_s^2 \tau /\epsilon$. In the lattice Boltzmann model, one must take into account second order corrections to the relaxation time, $\tau = \tau_{lb} - 1/2$, where $\tau_{lb}$ the lattice Boltzmann relaxation time.

From Eq.~\eqref{master:eq}, it is appreciated that the spatial derivatives of the source also introduce errors in the solution. Thus, by annihilating higher order contributions, we are also eliminating spurious effects due to the derivatives of the source term. If the source term is smooth, this derivatives are negligible, however, for stiffer source terms, one can use the present approach to add accuracy to the solution. 
In order to investigate this issue, we solve the potential for the following charge density:
\begin{equation}\label{eq:source}
 \rho(x) = \epsilon \sin \left (\frac{2\pi x}{L} l \right ) \quad ,
\end{equation}
where the integer $l$ controls the smoothness of the derivatives of the charge density, and $L$ is the simulated length, with $x \in [0, L)$. We use periodic boundary conditions for simplicity, and $\tau_{lb} = 1$ (numerical units).

\begin{figure}
  \centering
  \includegraphics[angle=270, width=0.5\columnwidth]{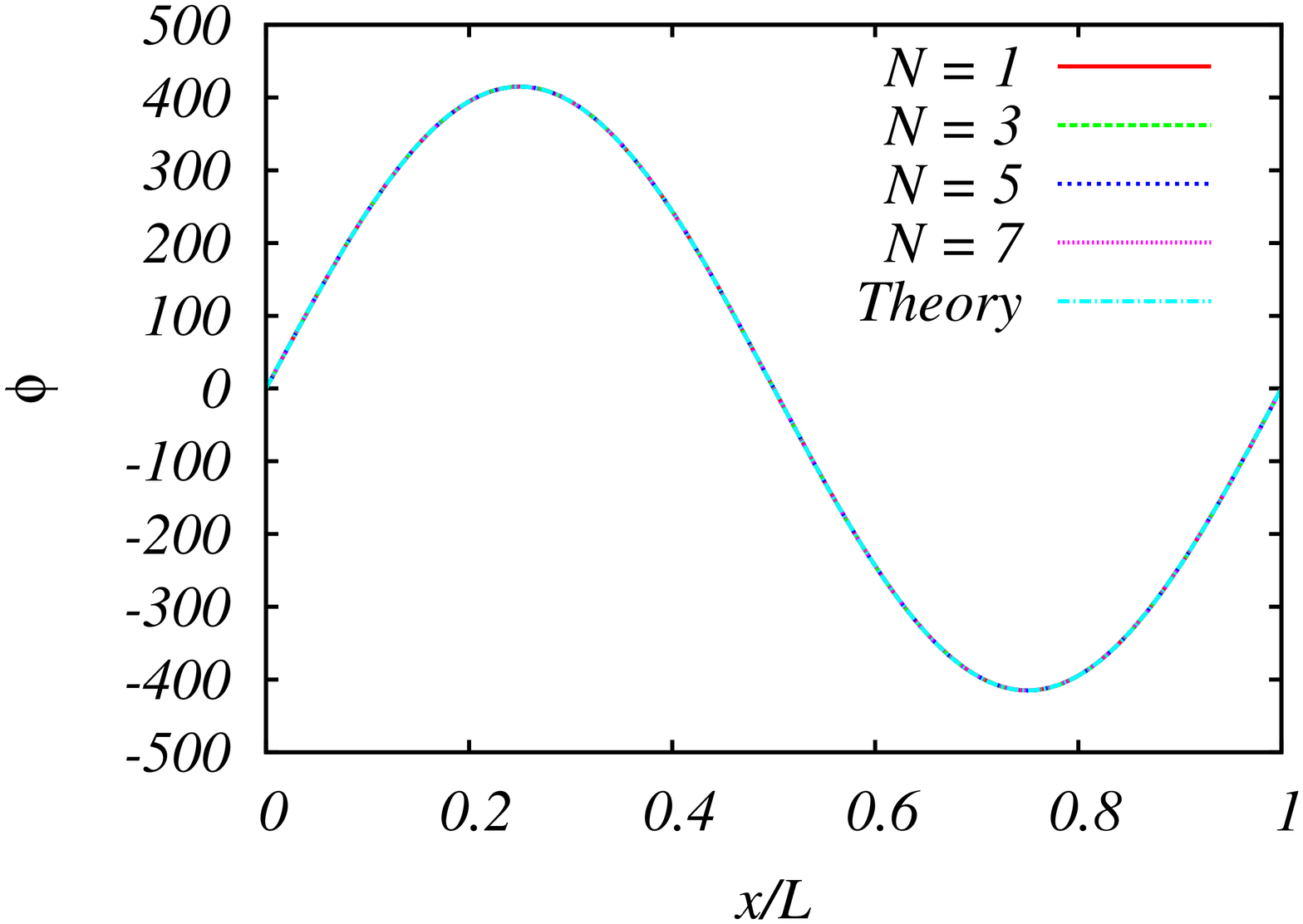}\includegraphics[angle=270, width=0.5\columnwidth]{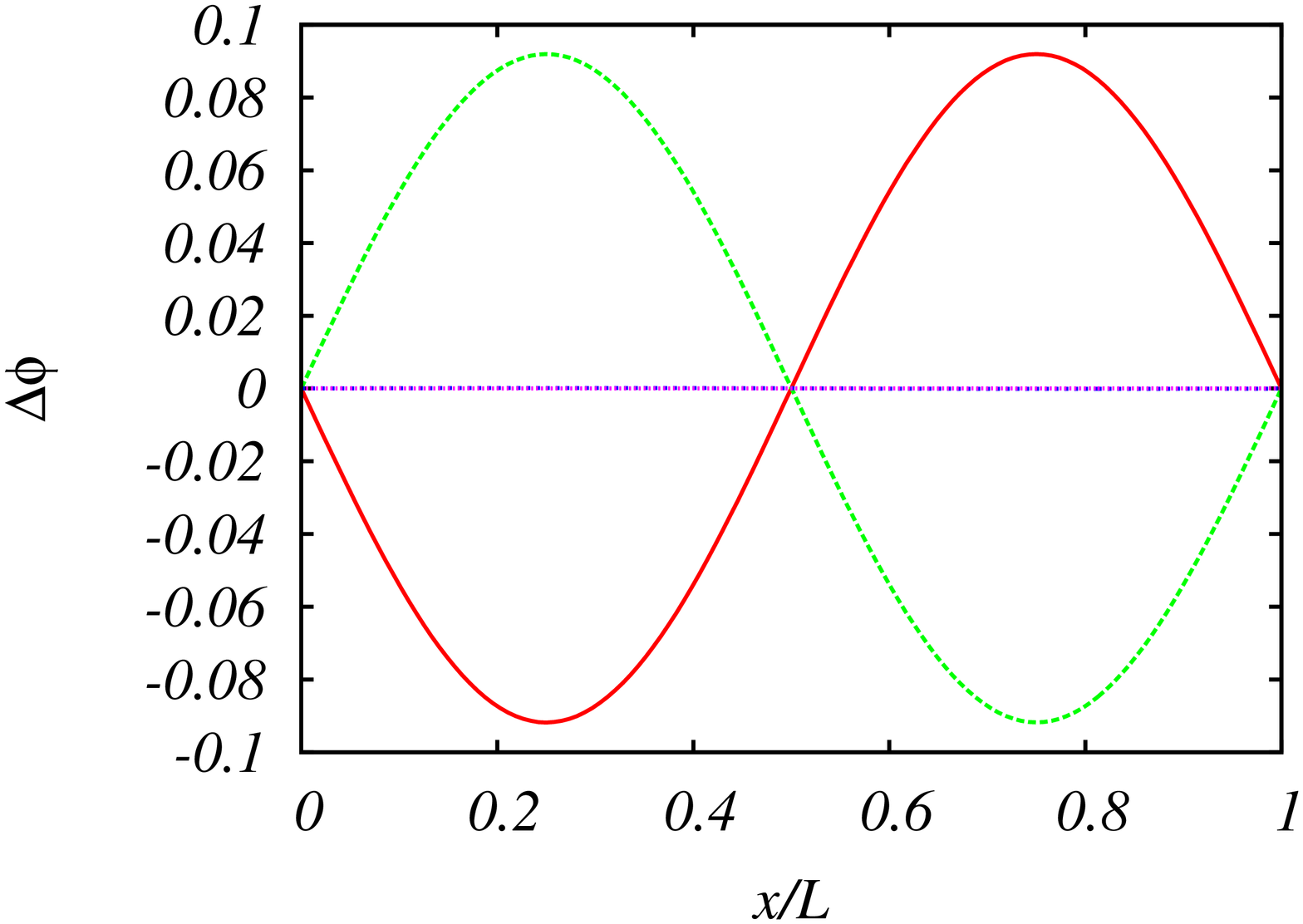}
  \caption{Solution of the Poisson equation for different cut-off $N$ and $l=1$. The absolute error is computed $\Delta \phi \equiv \phi_n - \phi_t$, where $\phi_n$ is the potential calculated with the lattice Boltzmann model, and $\phi_t$ is the analytical solution of Eq.~\eqref{poisson1}. The system size is $L = 128$ lattice cells.}\label{fig1}
\end{figure}
\begin{figure}
  \centering
  \includegraphics[angle=270, width=0.5\columnwidth]{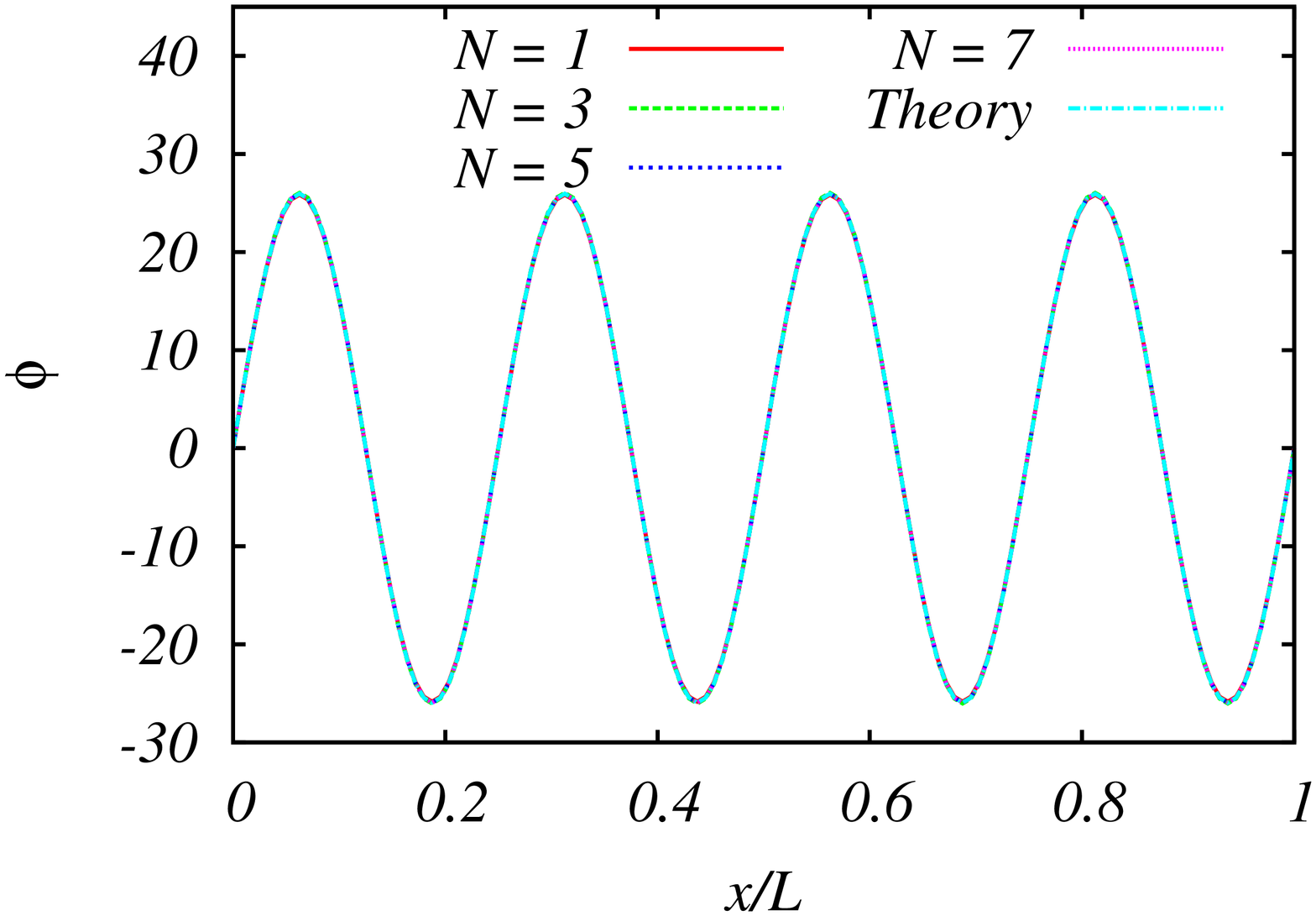}\includegraphics[angle=270, width=0.5\columnwidth]{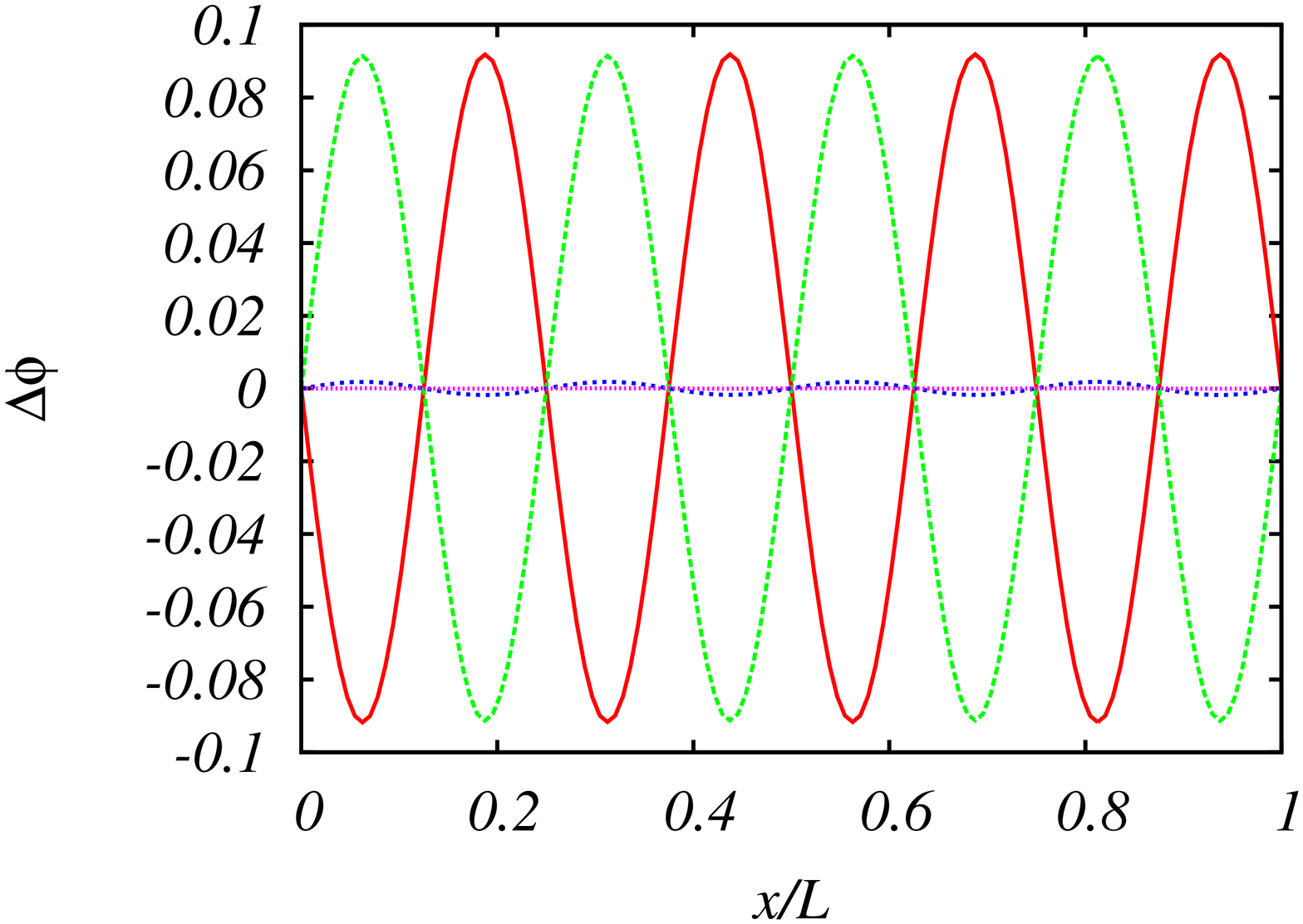}
  \caption{Solution of the Poisson equation for different cut-off $N$ and $l=4$. The absolute error is computed $\Delta \phi \equiv \phi_n - \phi_t$, where $\phi_n$ is the potential calculated with the lattice Boltzmann model, and $\phi_t$ is the analytical solution of Eq.~\eqref{poisson1}. The system size is $L = 128$ lattice cells.}\label{fig2}
\end{figure}
\begin{figure}
  \centering
  \includegraphics[angle=270, width=0.5\columnwidth]{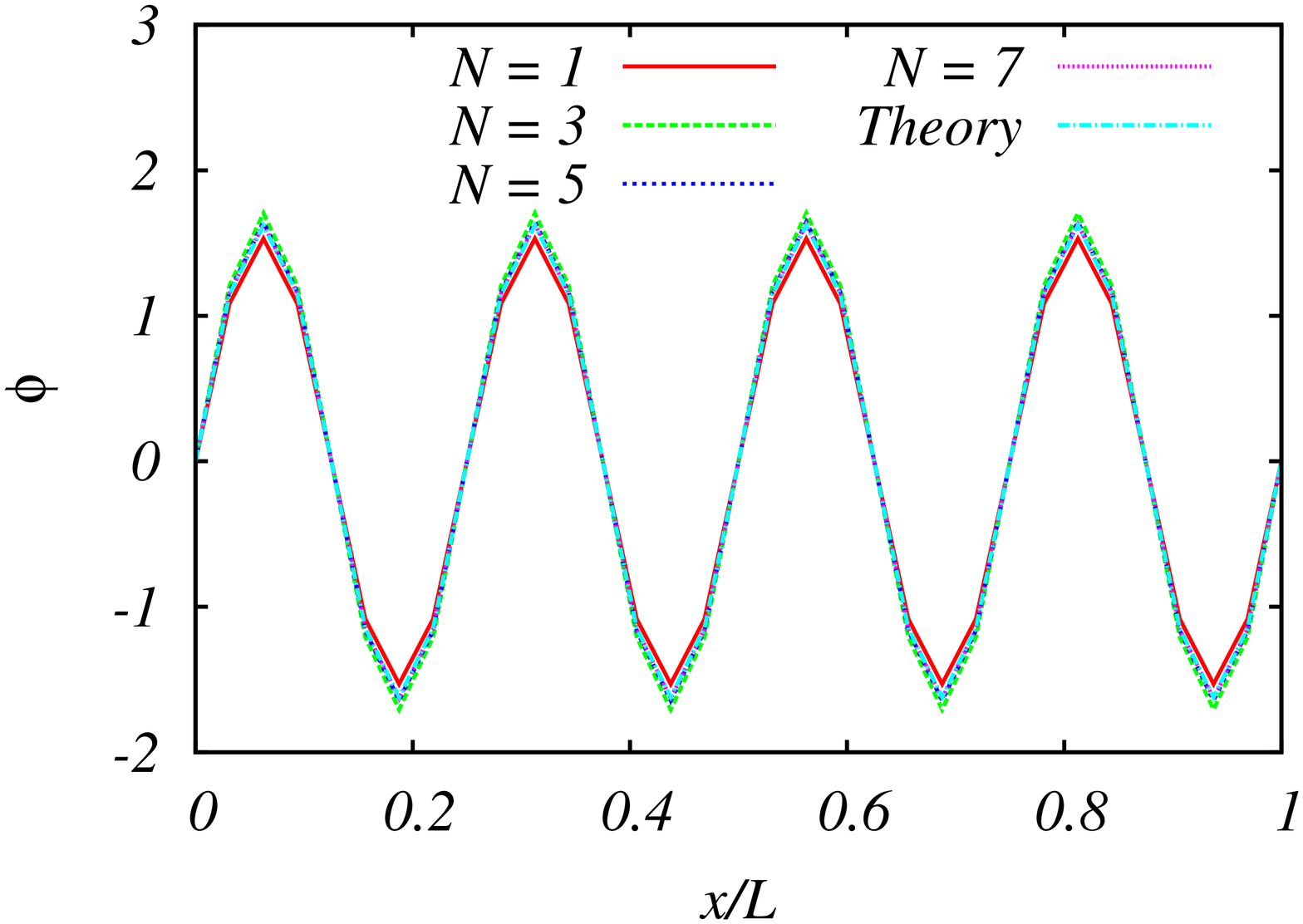}\includegraphics[angle=270, width=0.5\columnwidth]{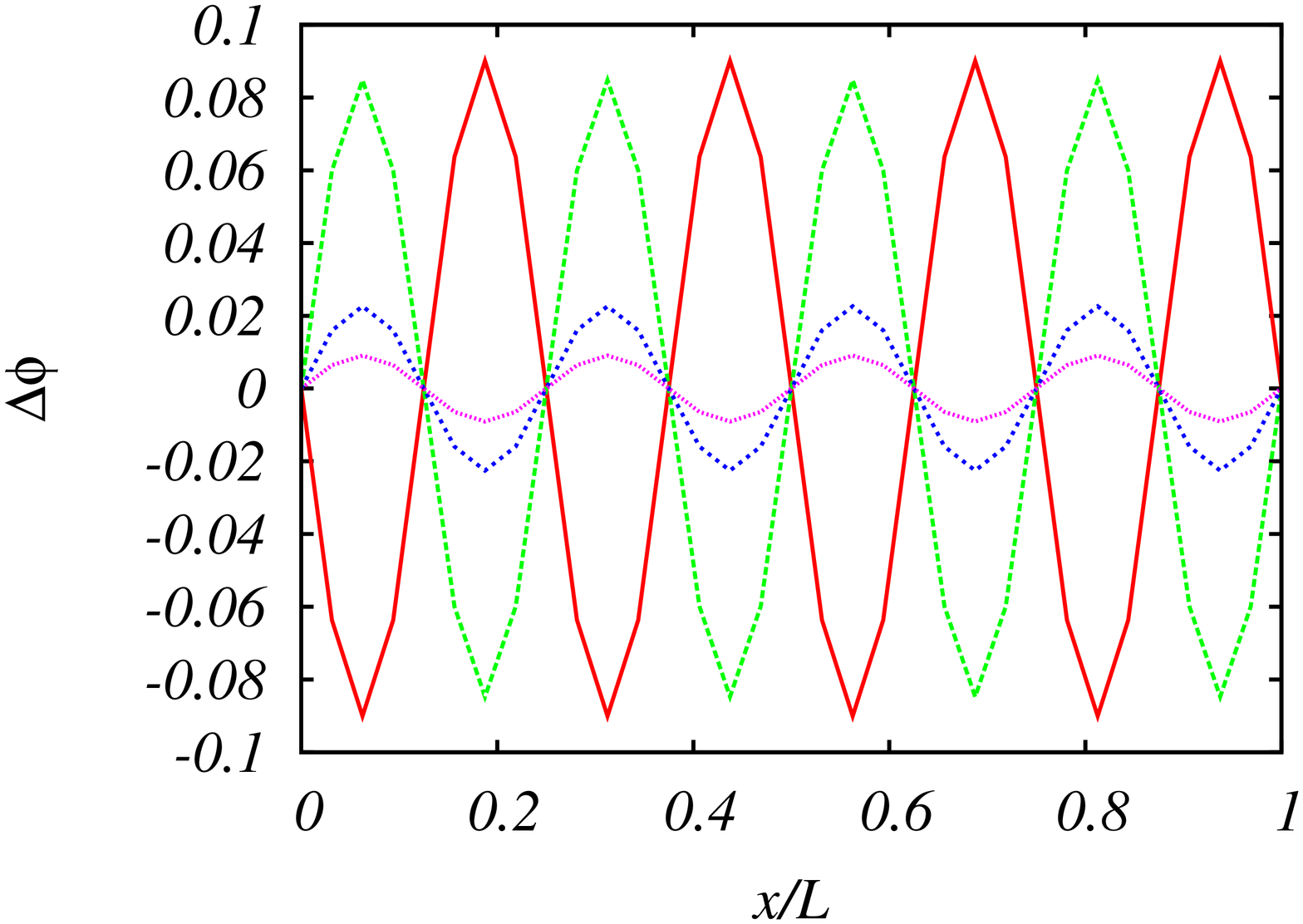}
  \caption{Solution of the Poisson equation for different cut-off $N$ and $l=4$. The absolute error is computed $\Delta \phi \equiv \phi_n - \phi_t$, where $\phi_n$ is the potential calculated with the lattice Boltzmann model, and $\phi_t$ is the analytical solution of Eq.~\eqref{poisson1}. The system size is $L = 32$ lattice cells.}\label{fig3}
\end{figure}
We express the relative error as
\begin{equation}
\label{ACCUR}
\epsilon = a L^{-\alpha}
\end{equation}
where the amplitude $a$ and scaling exponent $\alpha$ (order of accuracy) 
both depend on $N$.
From Fig.~\ref{fig1}, we see that for a relatively large system 
size, $L=128$, the first order expansion is sufficient to produce satisfactory results, with 
errors around $0.02 \%$ for $N=1$ and $N=3$, and below $10^{-3} \%$ for $N=5$ and $N=7$. 
In Fig.~\ref{fig2}, we see that by increasing the derivatives of the 
charge density (increasing $l$), the error increases for all values of the cut-off $N$. 
However, as we have mentioned before, the effect of removing higher-order errors
becomes crucial for the case of small system sizes. 
By decreasing the number of lattice cells (see Fig.~\ref{fig3}), the discrepancies 
become visual and the errors are around $5 \%$ for $N = 1, 3$, $1\%$ 
for $N = 5$, and less than $0.4 \%$ for $N = 7$.

\begin{figure}
  \centering
  \includegraphics[angle=270, width=0.5\columnwidth]{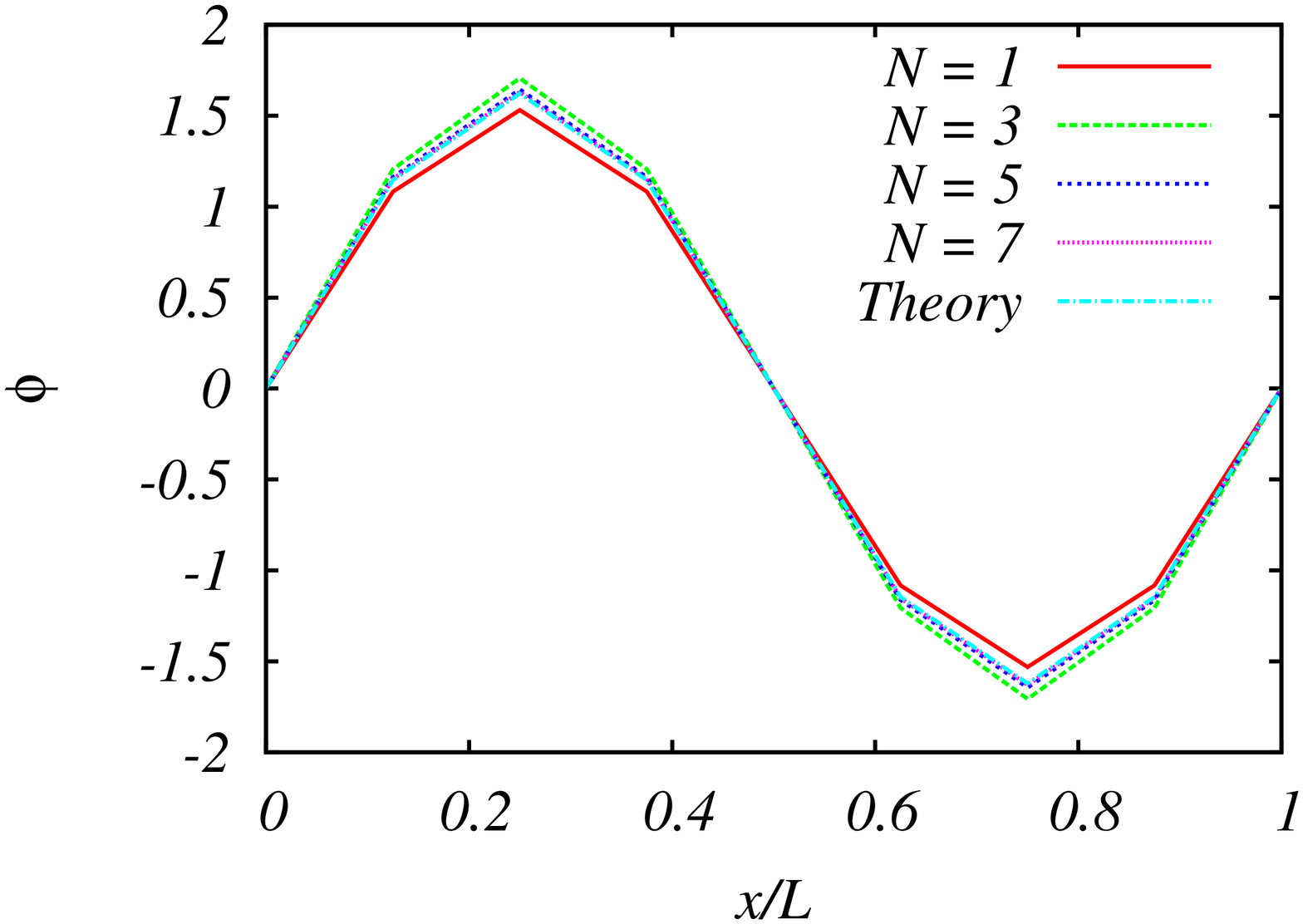}\includegraphics[angle=270, width=0.5\columnwidth]{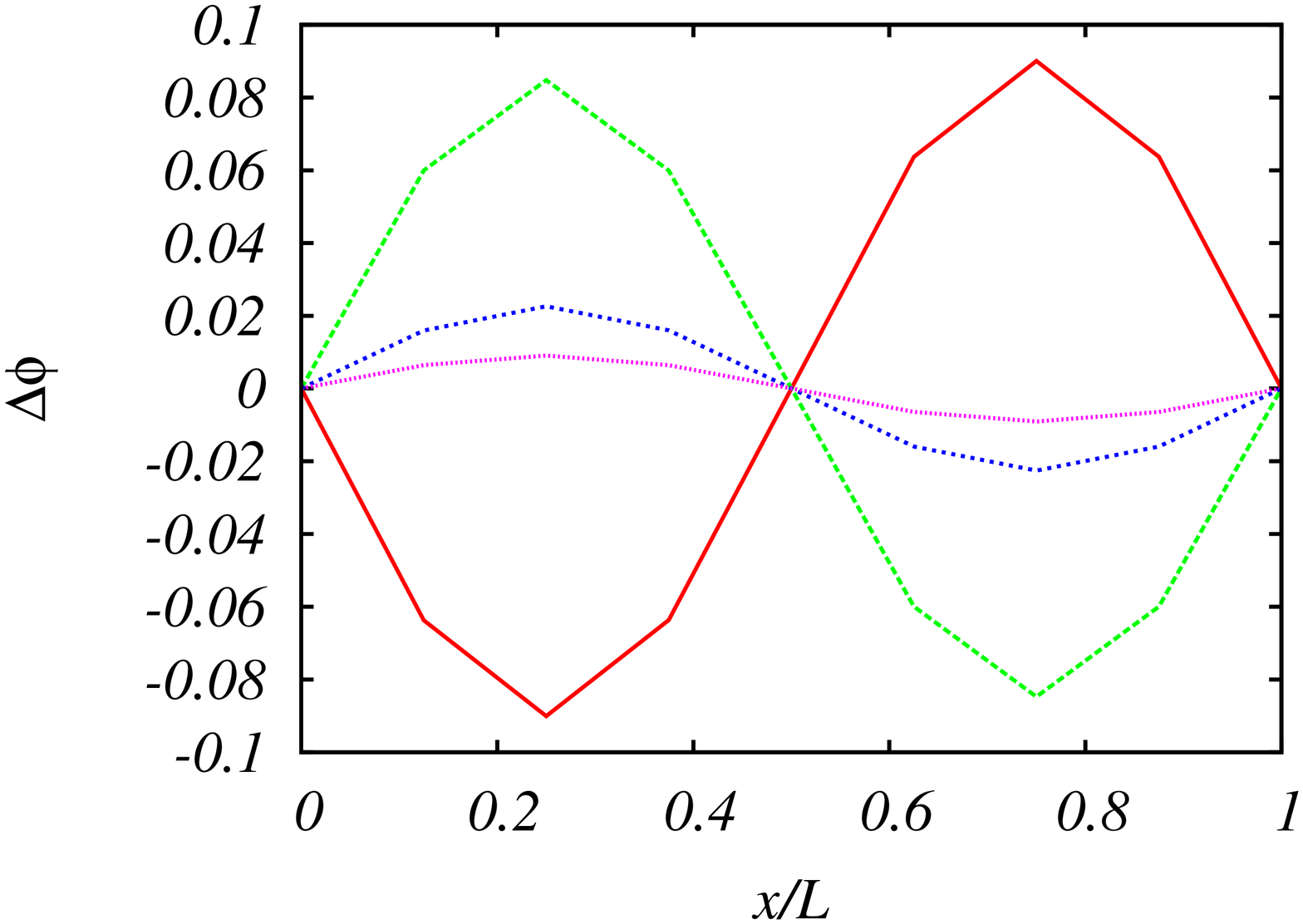}
  \caption{Solution of the Poisson equation for different cut-off $N$ and $l=1$. The absolute error is computed $\Delta \phi \equiv \phi_n - \phi_t$, where $\phi_n$ is the potential calculated with the lattice Boltzmann model, and $\phi_t$ is the analytical solution of Eq.~\eqref{poisson1}. The system size is $L = 8$ lattice cells.}\label{fig4}
\end{figure}

Finally, by further reducing the lattice resolution, $L = 8$ cells, and considering 
only one oscillation, $l=1$, we see that the errors for the first order expansion, $N=1$, are 
around $7 \%$, while for the highest values of $N = 7$ they remain below $0.5\%$. 
The fact that one can obtain very small errors by increasing the order of the expansion in the velocity space of the equilibrium distribution and source term, opens up the possibility of highly accurate solutions on very small grids. 

Increasing the order of the expansion inevitably implies an overhead 
of numerical operations per time step, thereby slowing down the simulations. 
On the other hand, higher order lattices also bring aditional side-benefits
such as a higher sound speed $c_s$, leading to a larger diffusion coefficient 
and, consequently, to a faster path to the steady-state solution. In order to highlight the concrete advantages of our approach, we next implement the same simulations for different orders $N$ and inspect the CPU time, number of iterations, and system size, required to achieve a specific level of accuracy. 

\subsection{Computational performance}

Let us next change the order of the expansion and analyse how the relative 
error decreases with the system size. 
\begin{figure}
  \centering
  \includegraphics[width=0.5\columnwidth]{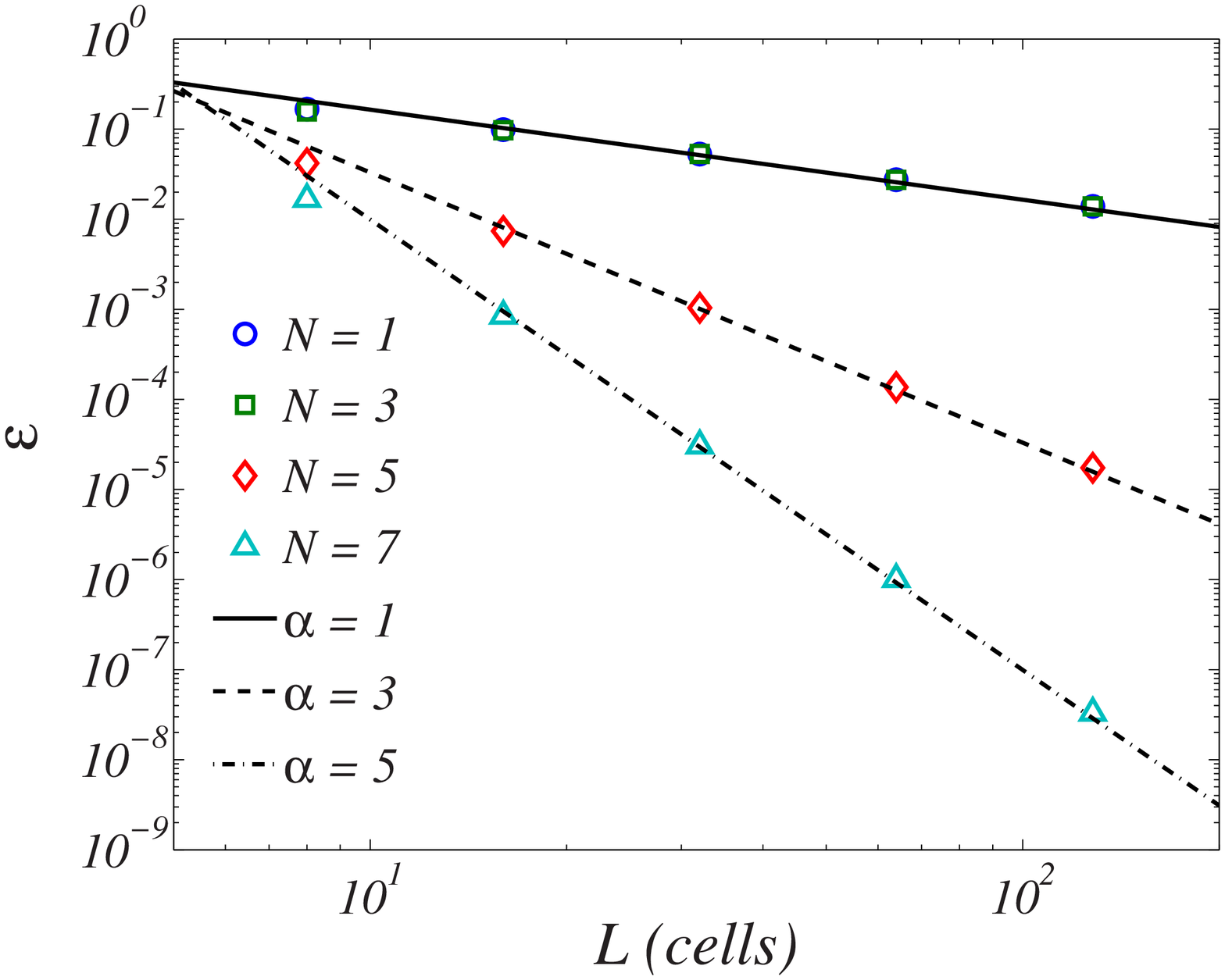}\includegraphics[width=0.5\columnwidth]{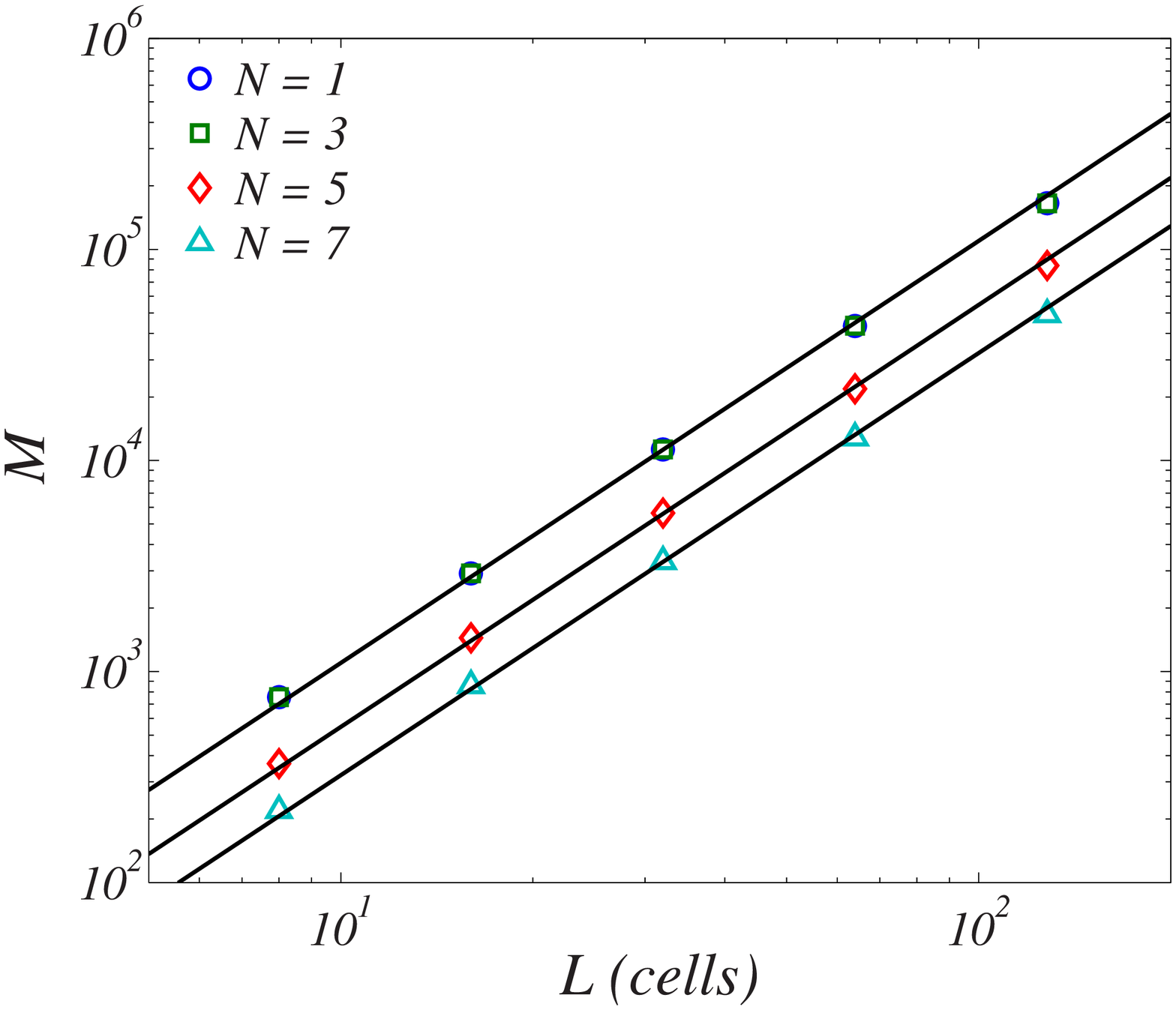}   
  \caption{(left) Relative error as a function of the system size $L$ for each cut-off value $N$ of the Hermite polynomials expansion. The lines denote the fitting curves using the general expression, $\epsilon = a_N L^{-\alpha}$ (right).
 Number of iterations as a function of the system size $L$ for each cut-off value of $N$. The lines denote the fitting curve using the general expression, $M = b_N L^2$, $M$ being
the number of time-steps.}\label{fig5}
\end{figure}
From Fig.~\ref{fig5} (left), we see that, according to our expectations, by increasing the order of the expansion, the order of convergence, measured by the exponent $\alpha$, also increases and in a very substantial way. From the same figure (right), we can also observe that for all values of $N$, the number of time-steps $M$ scales quadratically with the lattice size, and decreases with the order of the expansion $N$. The former feature is the typical signature of diffusion process, while the second reflects the fact that
higher order lattice deliver a larger sound speed, hence they take less 
time-steps to achieve a given level of accuracy.

The computational time required to achieve a given accuracy clearly depends
on the computational speed of the model on each given machine. 
This is conveniently expressed in terms of the number of lattice sites 
updates per CPU second, $\xi_N$. 
This quantity is expected to be a decreasing function of $N$, since higher-order 
quadratures require more operations per lattice site.

Actual measurements give the values reported in Table \ref{table1}. 
Although no clear trend can be extracted from these measurements, it is observed that
the cases $N=5$ and $N=7$ are about five times more expensive than the cases $N=1$ and $N=3$. 
Also to be noted that $\xi_7 > \xi_5$, which is counter-intuitive. 
However, upon inspecting the velocity vectors (see Appendix \ref{app}), one can appreciate that for $N=7$ the velocities are consecutive (no velocity gap), while for $N=5$ they are not. 
Since, memory access is faster for consecutive elements, we expect the case
$N=7$ to perform better than $N = 5$. More generally, the dependence on $N$ appears to be highly dependent on memory access patterns, which are not easily quantified through simple scaling relations.

The total CPU time, $t_{L,N}$, it takes to attain steady-state 
for a given system size $L$ can be written as follows:
\begin{equation}
\label{TL3}
t_{L,N} = \frac{b_N}{\xi_N} L^3 \quad .
\end{equation}
In the above, the cubic dependence derives from the $L^2$ diffusive scaling, as combined
with the number of sites $L$ (the space-time computational volume of a diffusive process
scales like $L^{2+D}$ in $D$ spatial dimensions).  
Finally, $b_N$ is a relative scale for the number of time-steps, which is expected to
decrease at increasing $N$ because of the increasing sound speed. The actual values of obtained in the simulations are reported in Table \ref{table1}. 
\begin{table}
  \tbl{Parameters $a_N$ and $b_N$  for each value of $N$. 
Also shown are the accuracy exponent $\alpha_N$ and
the number of lattice sites updated per second, $\xi_N$, in Msites/s (Million sites per second).}   
{  \begin{tabular}{|c|c|c|c|c|c|}\hline
    N & $a_N$ & $b_N$ & $\xi_N$ (Msites/s) & $\xi_N/b_N$ & $\alpha_N$\\ \hline
    1 & $1.645$ & $10.98$ & $26.2$ & $2.39$ & $1$ \\ \hline
    3 & $1.645$ & $10.98$ & $21.4$ & $1.95$ & $1$ \\ \hline 
    5 & $33.11$ & $5.46$ & $5.44$ & $0.997$ & $3$ \\ \hline
    7 & $992.3$ & $3.23$ & $5.76$ & $1.78$ & $5$ \\ \hline
  \end{tabular} \label{table1}}
\end{table}
As an example, for the case $L=100$, we find that the simulations 
take $0.42$ (65629 time steps), $0.51$ (65629 time steps), $1.0$ (33769 time steps), and $0.56$ (20079 time steps) seconds, for $N=1,3,5,7$ respectively. 

This shows that the total computational time remains within a factor two 
across all values of $N$, while the error with respect to the analytical solution 
decreases dramatically at increasing $N$ (see Fig.~\ref{fig5}).  
Hence, we can conclude that increasing $N$ leads to a dramatic boost of accuracy 
at a very moderate extra computational cost. Finally, we have inspected the computational time to steady-state for a given accuracy, at changing the size of the problem and the order of the quadrature.  By combining the previous relations, (\ref{ACCUR}) and (\ref{TL3}), we obtain:
\begin{equation}
\label{TEPS}
t_{N}(\epsilon) = \frac{b_N}{\xi_N}\left ( \frac{a_N}{\epsilon}\right )^{3/\alpha_N} \quad .
\end{equation}
The plot reports the computational time and system size needed to achieve 
a prescribed relative error, for the case in point $\epsilon = 10^{-3}$.  
\begin{figure}
  \centering
  \includegraphics[width=0.5\columnwidth]{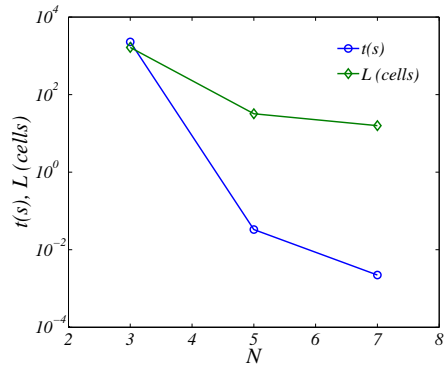}   
  \caption{Computational time $t_N$ and system size $L$ needed to recover the solution of the 
Poisson equation with a relative error of $0.1 \%$, for different $N$.}\label{fig6}
\end{figure}
From fig.~\ref{fig6}, we see that by increasing the order of the expansion in 
Hermite polynomials, it is possible to reduce the computational time by 
several orders of magnitude (up to six). 

These data refer specifically to the solution of the Poisson 
equation, but we have reasons to believe that similar conclusions 
would hold for other partial differential equations with a kinetic-theory background, 
including many linear non-linear partial differential equations
of great relevance in many branches of modern science. 

\subsubsection{Comparison with multi-grid methods}

In order to gain information about the performance of this approach, as compared 
with existing tools to solve the Poisson equation, we have implemented simulations 
using the multi-grid (MG) method \cite{multi1,multi2}. The number of nodes in the MG simulations is denoted by $L$,  like for the lattice Boltzmann method. Fixing $l=1$ (from Eq.~\eqref{eq:source}), in Table \ref{table2} we show the relative errors per node, $\epsilon/L$, together with the corresponding computational time. Note that for $N=3$, the MG method is faster than our approach, with the same second order accuracy. By increasing $N$, however, the MG becomes increasingly faster, but also increasingly less accurate  at a given grid size $L$. For instance, in order to attain the same accuracy of the case $N=7$ on $L=64$, the MG solver requires $L=16384$ grid points, being still faster by about a factor $3$. Given that $N=7$ requires $13$ arrays, the LB memory saving is about a factor $20$. These figures indicate that the present high-order LB Poisson solver is consistently slower than MG, however it also takes correspondingly less memory to achieve high levels of accuracy. 
Given that MG represents the state-of-the art for fast Poisson solvers, these can be
regarded as satisfactory results.

In summary, we have shown that the present  LB approach provides a viable option for the 
solution of Poisson equation, in the sense of simplicity and possibly also in terms
of memory demand at high accuracy.
\begin{table}
  \tbl{Average relative error per node, $\epsilon/L$, for the solution of the Poisson equation using different orders $N$ and the multi-grid method ($MG$). Inside the parenthesis, we report the computational time in milliseconds.}  
{  \begin{tabular}{|c|c|c|c|c|c|}\hline
    $L$ & $N=3$ & $N=5$ & $N=7$ & $MG$ \\ \hline
    16 & $10^{-2} (2)$ & $9\times 10^{-4} (2)$ & $10^{-4} (2)$ & $10^{-2} (< 1)$ \\ \hline
    32 & $3\times 10^{-3} (4)$ & $6\times 10^{-5} (4)$ & $2\times 10^{-6} (6)$ & $3\times 10^{-3} (< 1)$ \\ \hline 
    64 & $8\times 10^{-4} (26)$ & $4\times 10^{-6} (29)$ & $3\times 10^{-8} (38)$ & $8\times 10^{-4} (< 1)$ \\ \hline
    $2^{14}$ & $-$ & $-$ & $-$ & $2\times 10^{-8} (12)$ \\ \hline
  \end{tabular} \label{table2}}
\end{table}

\section{Conclusions}

Summarizing, we have studied the effect of higher order terms on the steady-state 
solution of the diffusion equation, using a lattice version of the 
Boltzmann equation with high-order (non-gaussian) equilibria. 
For the numerical solution, we have developed a higher order lattice Boltzmann 
model capable of reproducing up to seventh order moment of the 
equilibrium distribution and source charge. 
This permits to cancel exactly error terms up to seventh order. 
For the validation and study of our approach, we have solved the one-dimensional
Poisson equation, and found that the computational time to steady-state, at a prescribed
level of accuracy, can be reduced by up to six orders of magnitude as compared to standard
LB formulations.
The six orders gain in computational time is due to the fact that higher order lattices 
permit to utilize much smaller grid sizes.
Moreover, since they contain a large number of discrete velocities, they
allow faster propagation within the lattice, hence a correspondingly faster 
attainement of the steady-state.
We have also compared with state-of-the art multigrid solvers, and found that high-order 
LB is nearly competitive in efficiency at a lower memory demand.
 
Extensions to two and three-dimensions, as well as time-dependent diffusion
and advection-diffusion equations will make the object of future research.

\section*{Acknowledgements}
  We acknowledge financial support from the European Research Council (ERC) Advanced
  Grant 319968-FlowCCS. 

\appendix

\section{Quadrature of the Boltzmann Equation}\label{app}

Using Eq.~\eqref{quadrature} for each order of expansion $N$, we can calculate the velocity vectors $v_i$ with the respective weights $w_i$. In addition, each set of $w_i$ and $v_i$ provides a characteristic normalised speed $c_s$, $c_s^2 = \sum_{i=0}^M w_i v_i^2$. Here we only show the cases for $N = 3, 5$, and $7$, since the simulations for $N = 1$ were performed with the same lattice than for $N = 3$. For $N = 3$, we have
\begin{eqnarray}
\begin{aligned}
	w_0 &= 0.6366469031260781628443461\\
	w_1 &= 0.18141458774368577505004149\\
	w_2 &= 0.18141458774368577505004149\\
	w_3 &= 0.0002619606932751435277854615\\
	w_4 &= 0.0002619606932751435277854615\\
	v_0 &= 0, v_1=1, v_2=-1, v_3=3, v_4=-3\\
	c_s^2 &= 1 - \sqrt{2/5}\quad ,
\end{aligned}
\end{eqnarray}
for $N = 5$,
\begin{eqnarray}
\begin{aligned}
	w_0 &= 0.45813515550767658573\\
	w_1 &= 0.23734280857794043891\\
	w_2 &= 0.23734280857794043891\\
	w_3 &= 0.032324653788654934092\\
	w_4 &= 0.032324653788654934092\\
	w_5 &= 0.0012640621515365148385\\
	w_6 &= 0.0012640621515365148385\\
	w_7 &= 8.977280298192933351\times 10^{-7}\\
	w_8 &= 8.977280298192933351\times 10^{-7}\\
	v_0& = 0, v_1=1, v_2=-1, v_3=2, v_4=-2, \\
	v_5& = 3, v_6=-3, v_7=5, v_8=-5\\
	c_s^2 &= 0.86952909818721338884964163917209\quad ,
\end{aligned}
\end{eqnarray}
and finally, for $N = 7$,
\begin{eqnarray}
\begin{aligned}
	w_0 &= 0.3455934552621565\\
	w_1 &= 0.2374599218260301\\
	w_2 &= 0.2374599218260301\\
	w_3 &= 0.07705730993964580\\
	w_4 &= 0.07705730993964580\\
	w_5 &= 0.011801423732312036\\
	w_6 &= 0.011801423732312036\\
	w_7 &= 0.0008552466009513439\\
	w_8 &= 0.0008552466009513439\\
	w_9 &= 0.00002884934614927074\\
	w_{10} &= 0.00002884934614927074\\
	w_{11} &= 5.209238332209471\times 10^{-7}\\
	w_{12} &= 5.209238332209471\times 10^{-7}\\
	v_0&=0, v_1=1, v_2=-1, v_3=2, v_4=-2\\
	v_5&=3, v_6=-3, v_7=4, v_8=-4, v_9=5\\
	v_{10}&=-5, v_{11}=6, v_{12}=-6\\
	c_s^2 &= 1.1544053947399681272395977588380\quad .
\end{aligned}
\end{eqnarray}

For the three dimensional case, up to fifth order, we can find the velocity vectors $\vec{v}_i$ and weights $w_i$ by solving Eq.~\eqref{quad3}. The values are shown in Table \ref{vector2}.
\begin{table}
  \centering
\tbl{Velocity vectors $\vec{v}_i$ with their corresponding weights $w_i$ for $N = 5$. There are $111$ velocity vectors and $10$ different weights.}  
  {\begin{tabular}{|c|c|}\hline
    $w_i$& $\vec{v}_i$ \\ \hline
    $0.15014405$ & $(0,0,0)$  \\ \hline
    $0.02500399$ & $(\pm 1,0,0), (0, \pm 1,0), (0,0,\pm 1)$ \\ \hline
    $0.04505812$ & $(\pm 1,\pm 1,0), (0, \pm 1,\pm 1), (\pm 1,0,\pm 1)$ \\ \hline    
    $7.81706951\times 10^{-7}$ & $(\pm 3,\pm 3,\pm 3)$ \\ \hline   
    $9.79290909\times 10^{-6}$ & $(\pm 3,\pm 3,0), (\pm 3,0,\pm 3), (0,\pm 3,\pm 3)$ \\ \hline   
    $1.81207346\times 10^{-4}$ & $(\pm 3,\pm 1,0), (\pm 3,0,\pm 1), (0,\pm 1,\pm 3)$ \\
     &  $(\pm 1,\pm 3,0), (\pm 1,0,\pm 3), (0, \pm 3, \pm 1)$ \\ \hline
    $6.03109965\times 10^{-4}$ & $(\pm 2,\pm 2,0), (\pm 2,0,\pm 2), (0,\pm 2,\pm 2)$ \\ \hline   
    $3.49417975\times 10^{-3}$ & $(\pm 2,\pm 1,\pm 1), (\pm 1,\pm 1,\pm 2), (\pm 1,\pm 2,\pm 1)$ \\ \hline   
    $1.05490305\times 10^{-2}$ & $(\pm 2,0,0), (0, \pm 2,0), (0,0,\pm 2)$ \\ \hline     
    $4.50016852\times 10^{-5}$ & $(\pm 3,0,0), (0, \pm 3,0), (0,0,\pm 3)$ \\ \hline     
  \end{tabular}
  \label{vector2}}  
\end{table}


\end{document}